\documentclass[fleqn,twoside]{article}                                                 
\usepackage{espcrc2}                                                                   
\usepackage{graphicx}                                                                  
                                                          
\newcommand{\AmS}{{\protect\the\textfont2                                              
  A\kern-.1667em\lower.5ex\hbox{M}\kern-.125emS}}   
 
\newcommand{\ud}{\mathrm{d}}

\title{Classical and Quantum Ensembles via Multiresolution.\\II. Wigner Ensembles}
\author{Antonina~N.~Fedorova and Michael~G.~Zeitlin
\address{
IPME RAS, St.~Petersburg, 
V.O. Bolshoj pr., 61, 199178, Russia\\
E-mail: zeitlin@math.ipme.ru, anton@math.ipme.ru\\
http://www.ipme.ru/zeitlin.html,
http://www.ipme.nw.ru/zeitlin.html}}

\begin{document}

%%%%%%%%%%%%%%%%%%%%%%%%%%%%%%%%%%%%%%%%%%%%%%%%%%%%%%%%%%%%%%%%%%%%%%%%%%%
\thispagestyle{empty}

\begin{center}
\begin{tabular}{p{130mm}}

\begin{center}
{\bf\Large CLASSICAL AND QUANTUM ENSEMBLES}\\
\vspace{5mm}

{\bf\Large VIA MULTIRESOLUTION.} \\
\vspace{5mm}

{\bf\Large II. WIGNER ENSEMBLES}\\

\vspace{1cm}

{\bf\Large Antonina N. Fedorova, Michael G. Zeitlin}\\

\vspace{1cm}

{\bf\it
IPME RAS, St.~Petersburg,
V.O. Bolshoj pr., 61, 199178, Russia}\\
{\bf\large\it e-mail: zeitlin@math.ipme.ru}\\
{\bf\large\it e-mail: anton@math.ipme.ru}\\
{\bf\large\it http://www.ipme.ru/zeitlin.html}\\
{\bf\large\it http://www.ipme.nw.ru/zeitlin.html}
\end{center}

\vspace{1cm}
\begin{center}
\begin{tabular}{p{100mm}}
We present the application of the variational-wavelet analysis to                  
the analysis of quantum ensembles in Wigner framework. 
(Naive) deformation                     
quantization, the multiresolution 
representations and the variational approach are the key points. We        
construct the solutions of Wigner-like equations 
via the multiscale expansions in the generalized          
coherent states or high-localized nonlinear eigenmodes in the base of the          
compactly supported wavelets and the wavelet packets. We demonstrate the appearance of (stable)
localized patterns (waveletons) and consider entanglement 
and decoherence as possible applications.    
\end{tabular}
\end{center}

\vspace{20mm}

\begin{center}
{\large Presented at IX International Workshop on Advanced} \\
{\large Computing and Analysis Techniques in Physics Research}\\
{\large ACAT03, December, 2003, KEK, Tsukuba, Japan}

\vspace{5mm}

{\large Nuclear Instruments and Methods in Physics Research A, in press}
\end{center}
\end{tabular}
\end{center}
\newpage

\begin{abstract} 
We present the application of the variational-wavelet analysis to                  
the analysis of quantum ensembles in Wigner framework. 
(Naive) deformation                     
quantization, the multiresolution 
representations and the variational approach are the key points. We        
construct the solutions of Wigner-like equations 
via the multiscale expansions in the generalized          
coherent states or high-localized nonlinear eigenmodes in the base of the          
compactly supported wavelets and the wavelet packets. We demonstrate the appearance of (stable)
localized patterns (waveletons) and consider entanglement 
and decoherence as possible applications.    
\vspace{1pc}
\end{abstract} 
\maketitle

\section{WIGNER-LIKE EQUATIONS}

In this paper we consider the  applications of 
a nu\-me\-ri\-cal\--\-ana\-ly\-ti\-cal technique based on local nonlinear harmonic analysis
(wavelet analysis, generalized coherent states analysis) 
to  the description of quantum ensembles.
The corresponding class of individual Hamiltonians has the form
\begin{eqnarray}
\hat{H}(\hat{p},\hat{q})=\frac{\hat{p}^2}{2m}+U(\hat{p},\hat{q}),
\end{eqnarray}
where $U(\hat{p}, \hat{q})$ is an arbitrary polynomial 
function on $\hat{p}$, $\hat{q}$, 
and plays the key role in many areas of physics [1]. The particular cases, 
related to some physics models, are considered in [2].
Our goals are some attempt of classification and the explicit numerical-analytical constructions
of the existing quantum states in the wide class of models.
There is a hope on the understanding of relation between the structure of initial Hamiltonians and
the possible types of quantum states and the qualitative type of their behaviour.
Most important in many areas are:
localized states, chaotic-like or/and entangled patterns, localized (stable) patterns 
(waveletons).
Our starting point is the general point of view of a deformation quantization approach at least on
the Moyal/Weyl/Wigner level [1].
In the naive calculations we may use the simple formal representation for star product:
\begin{eqnarray}
* &\equiv&\exp \Big(\frac{i\hbar}{2}(\overleftarrow\partial_q\overrightarrow\partial_p-
   \overleftarrow\partial_p\overrightarrow\partial_q)\Big)
\end{eqnarray}
In this paper we consider the calculations of the Wigner functions
$W(p,q,t)$ (WF) corresponding
to the classical polynomial Hamiltonian $H(p,q,t)$ as the solution
of the Wigner equation [1]:
\begin{eqnarray}
i\hbar\frac{\partial}{\partial t}W = H * W - W * H
\end{eqnarray}
and related Wigner-like equations.
According to the Weyl transform, a quantum state (wave function or density 
operator $\rho$) corresponds
to the Wigner function, which is the analogue in some 
sense of classical phase-space distribution [1].
We consider the following form of differential equations for time-dependent WF, $W=W(p,q,t)$:
\begin{eqnarray}
W_t=\frac{2}{\hbar}\sin\Big[\frac{\hbar}{2}
(\partial^H_q\partial^W_p-\partial^H_p\partial^W_q)\Big]\cdot HW
\end{eqnarray}
In quantum statistics the ensemble properties are described by the density
operator
\begin{equation}
\rho(t)=\sum_i w_i|\Psi_i(t)><\Psi_i(t)|, \quad \sum_iw_i=1
\end{equation}
After Weyl transform we have the following
 de\-com\-position via partial Wigner
functions \\
$W_i(p,q,t)$ for the full ensemble Wigner function:
\begin{equation}
W(p,q,t)=\sum_iw_iW_i(p,q,t) 
\end{equation}
where the partial Wigner functions
{\setlength\arraycolsep{1pt}
\begin{eqnarray}
&&W_n(q,p,t)\equiv\frac{1}{2\pi\hbar}\int\ud\xi{\rm exp}\Big(-\frac{i}{\hbar}p\xi\Big)\\
&&\Psi^*_n(q-\frac{1}{2}\xi,t)\Psi_n(q+\frac{1}{2}\xi,t)\nonumber
\end{eqnarray}
}
are solutions of proper Wigner equations:
{\setlength\arraycolsep{1pt}
\begin{eqnarray}
&&\frac{\partial W_n}{\partial t}=-\frac{p}{m}\frac{\partial W_n}{\partial q}+\\
&&\sum^{\infty}_{\ell=0}\frac{(-1)^\ell(\hbar/2)^{2\ell}}{(2\ell+1)!}
\frac{\partial^{2\ell+1}U_n(q)}{\partial q^{2\ell+1}}
\frac{\partial^{2\ell+1}W_n}{\partial p^{2\ell+1}}\nonumber
\end{eqnarray}
}
The next case describes the important decoherence process, where
we have collective and environment subsystems with their own Hilbert spaces 
%\begin{equation}
$
\mathcal{H}=\mathcal{H}_c\otimes\mathcal{H}_e
$.
%\end{equation}
Analysis is based on Weyl transform of Lindblad master equation [1]:
{\setlength\arraycolsep{1pt}
\begin{eqnarray}
&&\dot{W}=\{H,W\}_{PB}+
\sum_{n\geq 1}\frac{\hbar^{2n}(-1)^n}{2^{2n}(2n+1)!}\\
&&\partial^{2n+1}_q U(q)\partial_p^{2n+1}W(q,p)+
2\gamma\partial_p pW+D\partial^2_pW\nonumber
\end{eqnarray}
}
In the next section we consider the variational-wavelet approach for the solution of all
these Wigner-like equations (3), (4), (8), (9) for the case of an 
arbitrary polynomial $U(q, p)$, which corresponds to a finite number 
of terms in the series expansion in (4), (8), (9) 
or to proper finite order of $\hbar$.
The localized bases/states are the natural generalization of standard coherent, 
squeezed, thermal squeezed states [1],
which correspond to quadratical systems (pure linear dynamics) with Gaussian Wigner functions.
Representation of underlying symmetry group (affine group in the simplest case) 
on the proper functional space of states generate the exact multiscale expansion
which allows to control contributions to
the final result from each scale of resolution from the whole underlying 
infinite scale of spaces. 
Numerical calculations according to methods of part I
explicitly demonstrate the quantum interference of
generalized localized states, pattern (entangled-like) formation from localized eigenmodes and 
the appearance of (stable) localized patterns (waveletons).

\section{VARIATIONAL MULTISCALE REPRESENTATIONS}

We obtain our multiscale/multiresolution representations for solutions of Wig\-ner-like equations
via a variational-wavelet approach. 
We represent the solutions as 
decomposition into localized eigenmodes (regarding action of affine group, i.e.
hidden symmetry of underlying functional space of states) 
related to the hidden underlying set of scales [3]:
{\setlength\arraycolsep{0pt} 
\begin{eqnarray}
&&W_n(t,q,p)=\displaystyle\bigoplus^\infty_{i=i_c}W^i_n(t,q,p)
\end{eqnarray}}
where value $i_c$ corresponds to the coarsest level of resolution
$c$ or to the internal scale with the number $c$ in the full multiresolution decomposition
of underlying functional space ($L^2$, e.g.) corresponding to problem under consideration:
$
%\begin{equation}
V_c\subset V_{c+1}\subset V_{c+2}\subset\dots
%\end{equation}
$
and $p=(p_1,p_2,\dots),\quad q=(q_1,q_2,\dots),\quad x_i=(p_1,q_1,\dots,p_i,q_i)$ 
are coordinates in phase space.
In the following we may consider as fixed as variable 
numbers of particles. The second case corresponds to quantum statistical ensemble 
(via ``wignerization'' procedure) and will be considered in details elsewhere.
We introduce the Fock-like space structure
\begin{eqnarray}
H=\bigoplus_i\bigotimes_n H^n_i
\end{eqnarray}
for the set of n-partial Wigner functions (states):
$$
W^i=\{W^i_0,W^i_1(x_1;t),\dots,
W^i_N(x_1,\dots,x_N;t),\dots\}
$$
where
$W_p(x_1,\dots, x_p;t)\in H^p$,
$H^0=C,\quad H^p=L^2(R^{6p})$ (or any different proper functional spa\-ce), 
with the natural Fock space like norm: 
\begin{eqnarray}
&&(W,W)=W^2_0+\\
&&\sum_{i}\int W^2_i(x_1,\dots,x_i;t)\prod^i_{\ell=1}\mu_\ell\nonumber
\end{eqnarray}
First of all we consider $W=W(t)$ as a function of time only,
$W\in L^2(R)$, via
multiresolution decomposition which naturally and efficiently introduces 
the infinite sequence of the underlying hidden scales [3].
We have the contribution to
the final result from each scale of resolution from the whole
infinite scale of spaces (11).
The closed subspace
$V_j (j\in {\bf Z})$ corresponds to  the level $j$ of resolution, 
or to the scale j
and satisfies
the following properties:
let $D_j$ be the orthonormal complement of $V_j$ with respect to $V_{j+1}$: 
$
V_{j+1}=V_j\bigoplus D_j.
$
Then we have the following decomposition:
\begin{eqnarray}
\{W(t)\}=\bigoplus_{-\infty<j<\infty} D_j 
=\overline{V_c\displaystyle\bigoplus^\infty_{j=0} D_j},
\end{eqnarray}
in case when $V_c$ is the coarsest scale of resolution.
The subgroup of translations generates a basis for the fixed scale number:
$
{\rm span}_{k\in Z}\{2^{j/2}\Psi(2^jt-k)\}=D_j.
$
The whole basis is generated by action of the full affine group:
\begin{eqnarray}
&&{\rm span}_{k\in Z, j\in Z}\{2^{j/2}\Psi(2^jt-k)\}=\\
&&{\rm span}_{k,j\in Z}\{\Psi_{j,k}\}
=\{W(t)\}\nonumber
\end{eqnarray}
One of the key points (the so called Fast Wavelet Transform, FWT) 
of wavelet analysis approach demonstrates that for a large class of
operators wavelets are good approximation for true eigenvectors and the corresponding 
matrices are almost diagonal. FWT gives  the maximum sparse form of operators
under consideration [3].
So, let us denote our (integral/differential) operator from equations under 
consideration  
as  $T$ ($L^2(R^n)\rightarrow L^2(R^n)$) and its kernel as $K$.
We have the following representation:
\begin{equation}
<Tf,g>=\int\int K(x,y)f(y)g(x)\ud x\ud y
\end{equation}
In case when $f$ and $g$ are wavelets $\varphi_{j,k}=2^{j/2}\varphi(2^jx-k)$
(15) provides the standard representation for operator $T$.
Let us consider multiresolution representation
$
\dots\subset V_2\subset V_1\subset V_0\subset V_{-1}
\subset V_{-2}\dots
$. 
The basis in each $V_j$ is 
$\varphi_{j,k}(x)$,
where indices $\ k, j$ represent translations and scaling 
respectively. 
Let $P_j: L^2(R^n)\rightarrow V_j$ $(j\in Z)$ be projection
operators on the subspace $V_j$ corresponding to level $j$ of resolution:
$
(P_jf)(x)=\sum_k<f,\varphi_{j,k}>\varphi_{j,k}(x).
$ 
Let
$Q_j=P_{j-1}-P_j$ be the projection operator on the subspace $D_j$ ($V_{j-1}=V_j\oplus D_j$) 
then
we have the following 
representation of operator T which takes into account contributions from
each level of resolution from different scales starting with the
coarsest and ending to the finest scales [3]:
\begin{equation}
T=\sum_{j\in Z}(Q_jTQ_j+Q_jTP_j+P_jTQ_j).
\end{equation}
We remember that this is a result of presence of affine group inside this
construction.
The non-standard form of operator representation [3] is a representation of
operator T as  a chain of triples
$T=\{A_j,B_j,\Gamma_j\}_{j\in Z}$, acting on the subspaces $V_j$ and
$D_j$:
$
 A_j: D_j\rightarrow D_j, B_j: V_j\rightarrow D_j,
\Gamma_j: D_j\rightarrow V_j,
$
where operators $\{A_j,B_j,\Gamma_j\}_{j\in Z}$ are defined
as
$A_j=Q_jTQ_j, \quad B_j=Q_jTP_j, \quad\Gamma_j=P_jTQ_j.$
The operator $T$ admits a recursive definition via
\begin{eqnarray}
T_j=
\left(\begin{array}{cc}
A_{j+1} & B_{j+1}\\
\Gamma_{j+1} & T_{j+1}
\end{array}\right),
\end{eqnarray}
where $T_j=P_jTP_j$ and $T_j$ acts on $ V_j: V_j\rightarrow V_j$.
So, it is possible to provide the following ``sparse'' action of operator $T_j$
on sufficiently smooth function $f$:
\begin{equation}
(T_j f)(x)=\sum_{k\in Z}\left(2^{-j}\sum_{\ell}r_\ell f_{j,k-\ell}\right)
\varphi_{j,k}(x),
\end{equation}
in the wavelet basis $\varphi_{j,k}(x)=2^{-j/2}\varphi(2^{-j}x-k)$ where
\begin{equation}
f_{j,k-1}=2^{-j/2}\int f(x)\varphi(2^{-j}x-k+\ell)\ud x
\end{equation}
are wavelet coefficients and $r_\ell$  
are roots of some additional linear system of equations related to
the ``type of localization'' [3].
So, we have simple linear para\-met\-rization of
matrix representation of  our operators in localized wavelet bases
and of the action of
this operator on arbitrary vector/state in proper functional space.
The variational approach [2] reduces the initial problem 
to the problem of solution 
of functional equations at the first stage and 
some algebraical problems at the second one.
So, the solution is parametrized by the solutions of two sets of 
reduced algebraical
problems, one is linear or nonlinear
(depending on the structure of the generic operator $L$ [2]) and the rest are linear
problems related to the computation of the coefficients of the Galerkin-like 
algebraic equations [3].
As a result the solution of the equations from Sec. 1 has the 
following mul\-ti\-sca\-le decomposition via 
nonlinear high\--lo\-ca\-li\-zed eigenmodes 
{\setlength\arraycolsep{0pt}
\begin{eqnarray}
&&W(t,x_1,x_2,\dots)=
\sum_{(i,j)\in Z^2}a_{ij}U^i\otimes V^j(t,x_1,\dots),\nonumber\\
&&V^j(t)=
V_N^{j,slow}(t)+\sum_{l\geq N}V^j_l(\omega_lt), \ \omega_l\sim 2^l, 
\end{eqnarray}
}
$$U^i(x_s)=
U_M^{i,slow}(x_s)+\sum_{m\geq M}U^i_m(k^{s}_mx_s), \ k^{s}_m\sim 2^m
$$
which corresponds to the full multiresolution expansion in all underlying time/space 
scales.
The formulae (20) give the expansion into a slow part
and fast oscillating parts for arbitrary $N, M$. We may move
from the coarse scales of resolution to the 
finest ones for obtaining more detailed information about the dynamical process.
In this way one obtains contributions to the full solution
from each scale of resolution or each time/space scale or from each nonlinear eigenmode.
It should be noted that such representations 
give the best possible localization
properties in the corresponding (phase)space/time coordinates. 
Formulae (20) do not use perturbation
techniques or linearization procedures.
Numerical calculations are based on compactly supported
wavelets and wavelet packets and on evaluation of the 
accuracy on 
the level $N$ of the corresponding cut-off of the full system 
regarding norm (12):
%\begin{equation}
$
\|W^{N+1}-W^{N}\|\leq\varepsilon
$.
%\end{equation}
\begin{figure}[ht]
\centering
\includegraphics*[width=55mm]{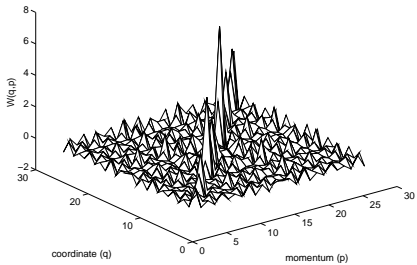}
%\centerline{\epsfxsize=3.9in\epsfbox{fz3.ps}}
\caption{Localized pattern-like (waveleton) Wigner function.}
\end{figure}
\begin{figure}[ht]
\centering
\includegraphics*[width=55mm]{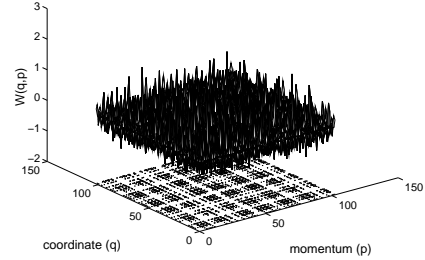}
%\centerline{\epsfxsize=3.6in\epsfbox{sm_0.01_512_5.ps}}
\caption{Entangled-like Wigner function.}
\end{figure}
Our (nonlinear) eigenmodes are more realistic for the modelling of 
nonlinear classical/quantum dynamical process  than the corresponding linear gaussian-like
coherent states. Here we mention only the best convergence properties of the expansions 
based on wavelet packets, which  realize the minimal Shannon entropy property
and the exponential control of convergence of expansions like (20) based on the norm (12).
Figures 1, 2 present the solutions, constructed
from the first 6 eigenmodes (6 levels in formula (20)), and demonstrate the stable localized 
pattern formation (waveleton) and complex 
entangled-like behaviour. Fig. 1 corresponds to (possible) result of superselection
(einselection) [1] after decoherence process started from entangled pattern demonstrated 
on Fig. 2.
It should be noted that
we can control the type of behaviour on the level of the reduced algebraical variational 
system [2].

\end{document}